\documentclass[a4paper,11pt]{article}
\pdfoutput=1 

\usepackage{jheppub} 

\usepackage[T1]{fontenc} 

\usepackage{color}

\def\be{\begin{equation}}
\def\ee{\end{equation}}
\def\bea{\begin{eqnarray}}
\def\eea{\end{eqnarray}}

\def\d{\delta}
\def\m{\mu}
\def\n{\nu}

\def\l{\lambda}

\def\f{\phi_{g}}

\def\bma{\begin{pmatrix}}
\def\ema{\end{pmatrix}}

\def\f{\phi}

\def\D{\nabla}

\def\bi{\begin{itemize}}
\def\ei{\end{itemize}}

\title{\boldmath Some Cosmological Consequences of Weyl Invariance}

\author[1]{Enrique Alvarez,\note{Corresponding author.}}
\author[] {Sergio Gonz\'alez-Mart\'{\i}n} 
\author[]{and Mario Herrero-Valea,}

\affiliation[]{Departamento de F\'{\i}sica Te\'orica and Instituto de F\'{\i}sica Te\'orica, IFT-UAM/CSIC\\Universidad Aut\'onoma, 20849 Madrid, Spain}

\emailAdd{enrique.alvarez@uam.es}
\emailAdd{mario.herrero@csic.es}
\emailAdd{sergio.gonzalez.martin@csic.es}

\abstract{We examine some Weyl invariant cosmological models in the framework of
generalized dilaton gravity, in which the action is made of a set of $N$ conformally coupled scalar fields. It will be shown that when the FRW ansatz for the spacetime metric is assumed, the Ward identity for conformal invariance guarantees that the gravitational equations hold whenever the scalar fields EM do so. It follows that any scale factor can solve the theory provided a non-trivial profile for a dilaton field. In particular, accelerated expansion is a natural
solution to the full set of equations.
}
\begin{document}
{\flushright{IFT-UAM/CSIC-15-011, ~~FTUAM-15-4}}

\maketitle
\flushbottom

\section{Introduction}
The purpose of the present paper is to explore some consequences of conformal invariance in Friedmann-Robertson-Walker (FRW) solutions in models inspired by the simplest dilaton gravity described by the action
\begin{align}\label{dilaton_action}
S_{D}=-\int d^{4}x\sqrt{|g|}\; \left(\frac{1}{12}R\f^{2}+\frac{1}{2}\D_{\m}\f\D^{\m}\f + \lambda \f^{4}\right)
\end{align}
where $\f$ is the dilaton field and $R$ the curvature scalar\footnote{Here we are using Landau's conventions for the Riemann and Ricci tensors.}. This action is invariant not only under diffeomorphisms but also under local scale (Weyl) transformations of the form
\begin{align}
&g_{\m\n}(x)\rightarrow \Omega^2(x) g_{\m\n}(x)\nonumber\\
&\phi(x)\rightarrow \Omega^{-1}(x) \phi(x)
\end{align}
which will act as a gauge symmetry.

The global sign in front of this action is irrelevant as it stands but it will matter if some other field is coupled to the metric. In particular, if one fixes the Weyl invariance by freezing the scalar field (which is something that is allowed classically although certainly not quantum mechanically)
\begin{align}
\f=\sqrt{12}\; M_p
\end{align}
then the action \eqref{dilaton_action} reduces to
\begin{align}\label{E-H}
S=-\int d^{4}x\sqrt{|g|}\; \left(M_{p}^2 R+ 144\lambda M_{p}^{4}\right)
\end{align}
giving rise to both an Einstein-Hilbert  and a cosmological constant terms. In that case, the global sign is relevant so far it induces the right sign for the gravitational coupling, which will be seen by any other possible field present in the action and coupled to \eqref{dilaton_action}.

Moreover, one could walk in the opposite direction and construct the dilaton gravity action by taking an averaging process over the Einstein-Hilbert lagrangian with cosmological constant term. In other words, one could start with \eqref{E-H} and perform a conformal transformation
\begin{align}
g_{\m\n}\rightarrow \frac{\f^{2}}{6 M_{p}^{2}}\;g_{\m\n}
\end{align}
making explicit the conformal mode of the metric to go from the Einstein frame to the Jordan frame, recovering \eqref{dilaton_action}.

In this way, one can see that the wrong sign in front of the kinetic energy of the scalar field is not a signal of unitarity violation but just a mechanism to compensate the absence of the conformal mode of the metric due to Weyl invariance. Indeed, the theory is unitary, since it is just General Relativity expressed in a diferent frame, and all on-shell UV divergences are just the same as in this theory\cite{AlvarezHVM}\cite{Hooft:2010ac}. Nevertheless, along this work we will stick to actions with no extra matter added, so the global sign will be irrelevant for our conclusions.

It is also possible to go further and add extra scalar fields to the action. A more general setting can be then described by the action
\begin{align}\label{action_general}
S=\int d^{4}x\sqrt{|g|}\; \left(\frac{1}{12}R N_{ab}\f^{a}\f^{b}+\frac{1}{2}N_{ab}\D_{\m}\f^{a}\D^{\m}\f^{b} + \left(\lambda_{ab}\f^{a}\f^{b}\right)^{2}\right)
\end{align}
where $N_{ab}$ is symmetric and takes the role of a metric in the configuration space spanned by the multiple scalar fields and $\lambda_{ab}$ is a matrix of interactions. This action is Weyl invariant in the same sense as \eqref{dilaton_action} if both $N_{ab}$ and $\lambda_{ab}$ are inert under conformal transformations. The Einstein-Hilbert term with right sign can then be obtained from this if at least one of the eigenvalues of $N_{ab}$ is negative, fixing then the corresponding eigenvector to be constant.

Models of this kind with just two scalar fields, as for instance
\begin{align}
S=\int d^{4}x\sqrt{|g|}\; \left(\frac{1}{12}R(\f_{1}^{2}-\f_{2}^{2})+\frac{1}{2}\D_{\m}\f_{1}\D^{\m}\f_{1}-\frac{1}{2}\D_{\m}\f_{1}\D^{\m}\f_{2}\right)
\end{align}
have been studied as a tool to obtain a transition from gravity to antigravity. The sign of the difference $\f_{1}^{2}-\f^{2}_{2}$ can depend on cosmic time, and can become positive; this then becomes an antigravity force, which yields an effect similar to a positive cosmological constant. Similar models have been considered by \cite{Bars} with different emphasis. This phase can be correlated with the matter energy density, in a manner similar to some tracking quintessence models. Gravity and antigravity are however separated by a strong coupling barrier where
\be
\phi_1=\pm \phi_2
\ee
which makes doubtful the possibility of travelling from one regime to the other.

Although these actions have been studied with focus in the previous considerations and also in the context of Higgs(-dilaton) inflation \cite{shap} outside the conformal point (in which the extra coupling constants in front of the non-minimal couplings can realize a large variety of inflationary scenarios), their Weyl invariant phase has not attracted much attention. Here we thus take an uncommon point of view and study \eqref{action_general} as a pure gravitational action in the conformal point, unvealing the consequences of Weyl (conformal) invariance for some simple cosmological models.

Conformal invariant theories are indeed not very intuitive. One of their most attractive facts is that no cosmological constant term is allowed in the lagrangian. This is due simply to the fact that the dimension zero operator
\begin{align}
{\cal O}\equiv M^{4}\int d^{4}x\sqrt{|g|}
\end{align}is not Weyl invariant.

There is no sense in which physical quantities can be  big or small (other than Weyl singlets); this concept is not conformally invariant. In particular, the (regularized) volume of spacetime can always be gauge fixed to one. 
\par
Conformal invariance has also dramatic implications for the equations of motion(EM). We would like in this paper to concentrate on those. To be specific, the Ward identity induced by this symmetry implies that
\begin{align}\label{Wards}
2g_{\m\n}\frac{\delta S_m}{\delta g_{\m\n}}+\sum_i \l_i \phi_i \frac{\delta S_m}{\delta \phi_i}=0
\end{align}
where $\phi_{i}$ refers to any other field content (whatever their spin) apart from the metric that could be in the action and $\lambda_{i}$ to their weight under Weyl transformations.

This spells out the fact that  the matter EM ensure the trace of the gravitational EM (which will be the full set in FRW spaces, since there is only one undetermined function in the metric). This is not modified by any interaction terms present, provided that they are conformal invariant.

In the following, we will study the consequences of Weyl invariance in the action \eqref{action_general} for Friedman-Robertson-Walker (FRW) solutions in the synchronous gauge, where the metric only has one single degree of freedom.

\section{Flat FRW solutions.}
The starting point is then the action \eqref{action_general}. The gauge invariant gravitational EM were computed in \cite{AlvarezHVM} and read
{\small \begin{align}
\nonumber &N_{ab}\f^{a}\f^{b}\left(\frac{1}{12}R_{\m\n}-\frac{1}{24}R g_{\m\n}\right)-\frac{1}{6}N_{ab}\left(\f^{a} \D_{\m}\D_{\n}\f^{b}-\f^{a}\D^{2}\f^{b} g_{\m\n}\right)+\\
&+\frac{1}{12}N_{ab}\left(4\D_{\m}\f^{a}\D_{\n}\f^{b}-\D_{\delta} \f^{a}\D^{\delta}\f^{b}g_{\m\n}\right)-\frac{1}{2}g_{\m\n}\left(\lambda_{ab}\f^{a}\f^{b}\right)^{2}=0
\end{align}}
\begin{align}
\frac{1}{12}RN_{ab}\f^{b}-\frac{1}{2}N_{ab}\D^{2}\f^{b}+2\lambda_{cd}\lambda_{ab}\f^{c}\f^{d}\f^{b}=0
\end{align}

In order to get some information in the simplest cosmological context it is enough to consider a simple ansatz
\be
ds^2= b^2 dt^2-a^2\d_{ij} dx^i dx^j \label{frw}
\ee
where both $a(t)$ and $b(t)$ depend on cosmic time only, being invariant under local Weyl transformations 
\bea
&&a\rightarrow \Omega a\nonumber\\
&& b\rightarrow \Omega b
\eea

There are many natural ways to fix the gauge for Weyl invariance here. In general, any gauge choice in the metric will reduce their degrees of freedom by relating the functions $a(t)$ and $b(t)$ in some manner. In the following, we choose the simplest {\em synchronous gauge} $b(t)=1$, where the expansion takes the form
\be
\theta\equiv \nabla_\m u^\m= 3{\dot{a}\over a}\equiv 3 H(t)
\ee
thus studying spatially flat FRW metrics\footnote{This case generalizes trivially to the other cases in which the three-dimensional spatial sections are of positive or negative constant curvature.}
\be
ds^2= dt^2-a^2\d_{ij} dx^i dx^j
\ee

In this case, the equations of motion are simple, only the diagonal elements of the gravitational EM survive. Let us denote $E_{\m\n}\equiv {\d S\over \d g^{\m\n}}$. Isotropy also implies that $E_{ij}=E \d_{ij}$, so we are left with three equations
\begin{align}
\label{eom}
&\frac{\delta S}{\delta \f^{a}}=\frac{N_{a}b}{2}\left[\ddot{\f}^{b}+\frac{\f^{b}}{a^2}\partial_{t}\left(a(t)\dot{a}(t)\right)+3H \dot{\f}^{b}\right]-2\lambda_{cd}\lambda_{ab}\f^{c}\f^{d}\f^{b}\\
&E_{00}=\frac{N_{ab}}{4}(H\f^{a}+\dot{\f}^{a})(H\f^{b}+\dot{\f}^{b})-\frac{1}{2}\left(\lambda_{ab}\f^{a}\f^{b}\right)^{2}\nonumber\\
&E=\frac{N_{ab}}{12}\left[\dot{a}^{2}\f^{a}\f^{b} +4a\dot{a}\f^{a}\dot{\f}^{b} -a^{2}\dot{\f}^{a}\dot{\f}^{b}+2a\ddot{a} \f^{a}\f^{b}+2a^{2}\f^{a}\ddot{\f}^{b}  \right]+\frac{a^2}{2}\left(\lambda_{ab}\f^{a}\f^{b}\right)^{2}\nonumber
\end{align}
where we have introduced the Hubble parameter $H=\frac{\dot{a}(t)}{a(t)}$ and a dot stands for derivative with respect to time. The scalar fields are taken to share the symmetries of the background, being thus isotropic.

However, not all of these equations are independent, since Bianchi identities have to be always satisfied. Actually, in any Diff invariant theory,
\be
\D_{\m}\left(\frac{\delta S}{\delta g_{\m\n}}\right)=0
\ee
which in the case of a FRW ansatz in four dimensions translates into
\be\label{Bianchi}
\partial_{t} E_{00}+3\frac{\dot{a}(t)}{a(t)}E_{00}+\dot{a}(t)a(t)E=0
\ee
implying that we can get rid of one of the EM, $E$ for instance, since it does not contain new information. Therefore Bianchi identities imply for the ansatz \eqref{frw} that there is only one independent gravitational equation.

On the other hand, as a consequence of Weyl invariance the following identity holds
\begin{align}\label{Ward}
2g^{\m\n}\frac{\delta S_D}{\delta g^{\m\n}}-\phi^{a} \frac{\delta S_D}{\delta \phi^{a}}=0
\end{align}
relating $E_{00}$ to a combination of all $\frac{\delta S}{\delta \f^{a}}$. Therefore, from $N+2$ equations, where $N$ is the number of scalar fields, we have go to just $N$ equations but still having $N+1$ undetermined functions, the $N$ scalar fields and the scale factor $a$.

These facts together then imply that the full set of gravitational equations is redundant. Any profile $a(t)$ yields a consistent solution of the full dilaton gravity EM, provided that the EM for the scalar fields are solved in this background. 

Therefore, the action \eqref{action_general} admits solutions in which the scale parameter can be set by hand without any other bound than it being regular enough for the EM of the scalar fields to be solved.
\section{A single scalar field}
As an example of what we have argued in the general case, let us take the simplest action \eqref{dilaton_action}, in which the metric $N_{ab}$ has only one component
\begin{align}
S_{D}=-\int d^{4}x\sqrt{|g|}\; \left(\frac{1}{12}R\f^{2}+\frac{1}{2}\D_{\m}\f\D^{\m}\f + \lambda \f^{4}\right)
\end{align}

In this case, the equations of motion also simplify and we just have
\begin{align}
&\frac{\delta S}{\delta \f}=\frac{1}{2}\left[\ddot{\f}+\frac{\f}{a(t)^2}\partial_{t}\left(a(t)\dot{a}(t)\right)+3H \dot{\f}\right]-2\lambda \f^{3}\\
\nonumber &E_{00}=\frac{1}{4}\left[ \f^{2}\left(\frac{\dot{\f}}{\f}+H\right)^{2}-2\lambda \f^{4}\right]\\
&E=\frac{1}{12}\left[-\f^{2}\dot{a}(t)^{2}-2a(t)\f\left(2\dot{a}(t)\dot{\f}+\f\ddot{a}(t)\right)+a(t)^{2}\left(\dot{\f}^{2}-2\f\ddot{\f}+6\lambda \f^{4}\right)\right]\nonumber
\end{align}

If we take here the spatial gravitational equation and substitute the value of $\ddot{\f}$ given by the EM for the scalar field, we find that it simplifies to
\begin{align}
-\frac{1}{12}\left[ \f^{2}\left(\frac{\dot{\f}}{\f}+H\right)^{2}-2\lambda \f^{4}\right]=0
\end{align}
which is exactly $-\frac{1}{3}$ of $E_{00}$, showing that, indeed and as it was expected due to Bianchi identities in the form \eqref{Bianchi}, there is in principle only one independent gravitational equation of motion.

Moreover, if we now construct the trace of the gravitational equations of motion
\begin{align}
g^{\m\n}\frac{\delta S}{\delta g^{\m\n}}=\frac{\f}{2}\left[\ddot{\f}+\frac{\f}{a(t)^2}\partial_{t}\left(a(t)\dot{a}(t)\right)+3H \dot{\f}\right]-2\lambda \f^{4}
\end{align}
we see that we recover exactly the EM of the scalar field multiplied by the field itself, as a simple check that the Noether identity for Weyl invariance \eqref{Ward} holds
\begin{align}
g^{\m\n}\frac{\delta S}{\delta g^{\m\n}}-\f\frac{\delta S}{\delta \f}=0
\end{align}

Therefore, we are left with just one independent equation, which we choose to be $E_{00}$ for convenience, but with two unknown functions, meaning that we can choose one of them freely. If we choose it to be the scale factor, then for any choice of a flat FRW metric, the profile for the scalar field can be trivially obtained by using the $E_{00}$ equation of motion
\begin{align}
\left(\frac{\dot{\f}}{\f}+H\right)^{2}=2\lambda \f^{2}
\end{align}
which yields straightforwardly the solution for the scalar field
\begin{align}
\f=e^{-q(t)}\left(C\pm\sqrt{2\lambda}\;\int_{1}^{t}dt_{1} e^{-q(t_{1})}\right)^{-1}
\end{align}
provided $\lambda\geq 0$, $C$ being an integration constant and with the function $q(t)$ defined by
\begin{align}
q(t)=\int_{1}^{t} du H(u)
\end{align}

As long as the Hubble parameter of a given metric is integrable, this metric is a solution of the dilaton gravity theory \eqref{dilaton_action}. This includes deSitter and anti-deSitter spaces as well as some other more exotic examples. Even accelerated expanding universes fit into this scheme without the need of adding any exotic component of matter as a quintessence\cite{Zlatev:1998tr}. Physically, this result may be interpreted as if the dilaton field backreacts to the configuration of spacetime, sitting in a configuration just in the proper way to substain any desired geometry the user wants.

\section{Conclusions.}
Some Weyl invariant cosmological models are examined in the context of dilaton gravity. It has been shown that when the FRW ansatz for the spacetime metric is assumed, the Ward identity for conformal invariance guarantees that the gravitational equations hold whenever the matter EM do so. Several surprising facts follow from this. For example, maximally symmetric spacetimes (de Sitter and anti de Sitter) are exact solutions even though the cosmological constant vanishes. Actually, no cosmological constant is allowed by conformal invariance.
\par
However, this is a consequence of  the fact that FRW ansatz only has one degree of freedom. Were more complicated  spacetime ansatze  considered, then only the trace of the gravitational equations would be implied by the matter EM.
\par
Let us finally note that all the simplest conformal invariants (which are expected to be the only physical observables of the theory\cite{Carrasco:2013hua}) vanish in our case, owing to the fact that FRW geometries are of Petrov type 0 (vanishing Weyl tensor), not showing then any singularity. Also the action $S_D$ itself vanishes, as it does for all solutions.
 \par
It is obvious that the low energy effective theory is not conformally invariant. A mechanism for the spontaneous breaking of Weyl invariance, connecting the invariant phase with a more realistic one, is needed before any phenomenological models can be built along the lines of the present paper.

\section*{Acknowledgments}
 This work has been partially supported by the European Union FP7  ITN INVISIBLES (Marie Curie Actions, PITN- GA-2011- 289442)and (HPRN-CT-200-00148) as well as by FPA2009-09017 (DGI del MCyT, Spain), FPA2011-24568 (MICINN, Spain) and S2009ESP-1473 (CA Madrid).  The authors acknowledge the support of the Spanish MINECO {\em Centro de Excelencia Severo Ochoa} Programme under grant  SEV-2012-0249. 
\appendix

\end{document}